\documentclass[12pt,fleqn]{article}

	\RequirePackage[latin1]{inputenc}  
\usepackage{times}
	\usepackage[english]{babel}	
		\usepackage{amssymb}
	\usepackage{vmargin}
	\setpapersize{A4}
	\setmarginsrb{2cm}{2.5cm}{2cm}{1.5cm}{0cm}{0cm}{0cm}{0cm}
	\usepackage{graphicx}
\renewcommand {\d} {\textrm{d}}
\newcommand {\eps}{\varepsilon}
\newcommand {\epse}{\varepsilon^{\textrm{\small e}}}
\newcommand {\epss}{\varepsilon ^{\textrm{\small s}}}
\newcommand {\epsnn}{\varepsilon_{\textrm{\small nn}}}

\newcommand {\C} {\mathcal{C}}
\newcommand {\sst} {\sigma^{\textrm{\small t}}}
\newcommand {\ssc} {\sigma^{\textrm{\small c}}}
\newcommand {\asst} {\alpha^{\textrm{\small t}}}
\newcommand {\assc} {\alpha ^{\textrm{\small c}}}

\begin{document}

\title{\textsf{\textbf{\large Behavior of cracked materials}}}
\author{\textsf{\large Marc Fran\c cois}}
\date{}
\maketitle

\begin{center}
\textsf{\small LMT Cachan - Université Pierre et Marie Curie,\\
61 av. du Président Wilson, 94235 Cachan CEDEX, France\\
francois@lmt.ens-cachan.fr}
\end{center}
\normalsize

\thispagestyle{empty}


\noindent\textbf{Abstract.} Due to their microstructure, quasi brittle materials present rough cracks. Under sliding of the crack lips, this roughness involves in one hand induced opening and in the other hand some apparent plasticity which is due to the interlocking of the crack lips combined with Coulomb's friction. The proposed model is written under the irreversible thermodynamics framework. Micromechanics uses the Del Piero and Owen's structured deformation theory. Opening of the crack depends upon the crack shape and the relative sliding of the crack lips. The thermodynamic force associated to the sliding has the mechanical meaning of the force acting in order to make the crack slide. Yield surface is defined as a limitation of this force with respect to the Coulomb's friction and the Barenblatt cohesion. The crack orientation is defined as the one for which the criterion is reached for the lowest stress level. A decreasing cohesion, respect to sliding is supposed. Tension and compression reference cases are envisaged.\\

\noindent\textbf{Introduction}\\

This work is in continuity with \cite{francois_05}, in which the problem of sliding rough fracture is treated is the case of compressive and shearing testing, relevant  to the behavior of civil engineering structure response under earthquake. The goal was to extend this approach to cases of tension and compression, predict the crack's orientation and the global response of the specimen.\\

\noindent\textbf{Micromechanics}\\
\begin{figure}[htbp]
\begin{center}
\includegraphics[scale=0.333]{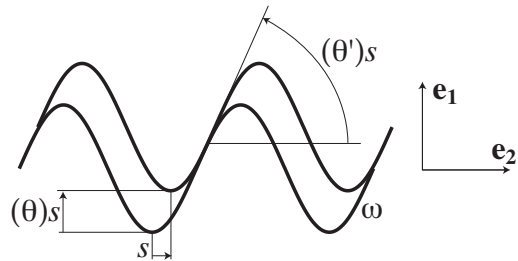}
\caption{sketch of the micromechanics}
\label{sketch}
\end{center}
\end{figure}
In case of periodic fracture, a micro-mechanical structured deformation analysis \cite{delpiero_93, francois_05} lead to a partition of the strain $\eps$ between it's elastic (reversible) $\epse$ and due to sliding (dissipative) $\epss$ parts (Eq. \ref{partoche}). The sliding of the crack is defined by a variable $s$ and the opening respect to the function $\theta(s)$. Considering the orthonormal basis $[\mathbf{e}_{1},\mathbf{e}_{2}]$, where $\mathbf{e}_{2}$ defines the normal to the (average) crack plane and $\mathbf{e}_{1}$ the sliding direction.
\begin{eqnarray}
\eps &=& \epse+\epss \label{partoche}\\
\epss &=& \theta(s) \left( \mathbf{e}_{1} \otimes \mathbf{e}_{1} \right) + s \theta(s) \frac{1}{2} \left( \mathbf{e}_{1} \otimes \mathbf{e}_{2}+\mathbf{e}_{2} \otimes \mathbf{e}_{1} \right) \label{defostruct}
\end{eqnarray}

The micro-mechanical point of view is sketched in the Fig. \ref{sketch}. The sliding $s$ induces the opening of the crack, respect to the opening function $\theta(s)$. This one depends upon the crack's shape $a\omega$ ($a$ is a physical length). In case of sinusoidal crack shape, naming $\xi$ the (metric) abscissa of the crack along $\mathbf{e}_{2}$, $A$ and $p$ the mathematical amplitude and periodicity, we have ($a$ vanishes in $\theta$):

\begin{equation}
\omega\left(\frac{\xi}{a}\right) = -A\cos\left(2\pi\frac{\xi}{ap}\right), \quad \theta(s)=A\left|\sin\left(\frac{\pi s}{p}\right)\right|\label{theta}
\end{equation}

\noindent\textbf{Thermodynamics: state}\\

The free energy $\rho \Psi$ is related to the elastic energy stored by the sound material (with $\C$ the Hook stiffness tensor). The derivative of it respect to state variables lead to the thermodynamic forces: the stress  $\sigma$ and the force $S$ (Eq. \ref{forceS}), that has the clear meaning of the projection of the stress $\sigma$ that have the will make the crack slide.
\begin{eqnarray}
2 \rho \Psi (\eps,s) &=& (\eps - \epss):\C:(\eps - \epss)\\
\sigma &=& \C:\eps, \quad 
S = \theta'(s) \sigma_{11}+\sigma_{12}\label{forceS}
\end{eqnarray}

\noindent\textbf{Elasticity domain for a crack}\\

\begin{figure}[htbp]
\begin{center}
\includegraphics{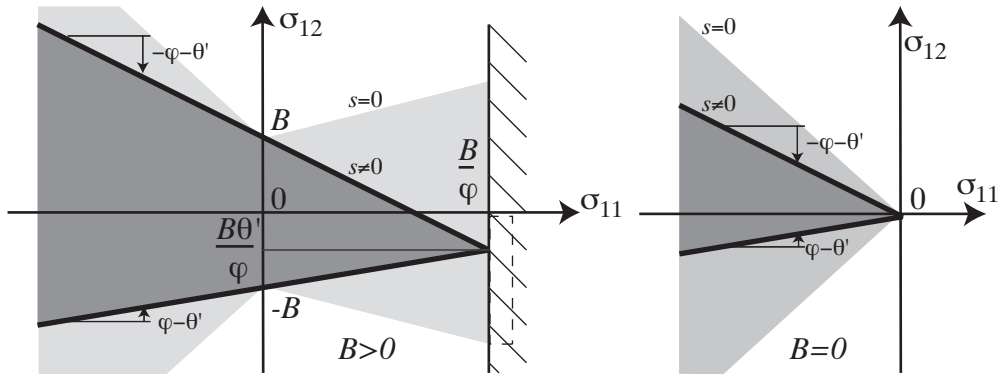}
\caption{Yield surface (active crack), for any $s$ and for $s=0$ (light gray)}
\label{yieldsurf}
\end{center}
\end{figure}

The yield surface $f(\sigma,S)$ is defined (in the thermodynamic forces space) as a limitation of the force $S$ respect to two phenomena: one cohesive and the other frictional. The fist, is a Barenblatt adhesive term $B_{0}\geqslant 0$ that represents cohesion, \emph{i.e.} when crack does not exist; this cohesion can be seen as the cement paste role in a concrete. The second member introduces the term $\varphi\geqslant 0$ that can be seen as a Coulomb's friction (at the macro scale level).

\begin{equation}
f(\sigma,S) = | S | - B_{0} + \varphi \sigma_{11}\label{f}
\end{equation}

The elasticity domain is then defined as $f<0$ and the dissipative evolution (siding of the fracture) occurs when $f=0$. Using Eq. (\ref{forceS}), the combination of them, $f\leqslant 0$, define the yield function (Eq. \ref{fcoo}) and the admissibility condition (Eq. \ref{admcoo}), in the $\mathbf{e}_{i}$ coordinates:
\begin{eqnarray}
|\theta'(s) \sigma_{11}+\sigma_{12}| &\leqslant& B_{0}-\varphi \sigma_{11}\label{fcoo}\\
\sigma_{11}&\leqslant& \frac{B_{0}}{\varphi}\label{admcoo}
\end{eqnarray}

The elasticity domain is represented onto Fig. \ref{yieldsurf}. Non admissible domain is hatched. It can be seen as the case when tension becomes large enough in order to separate crack's faces. When $B_{0}=0$, the yield surface tends to a Mohr-Coulomb's one, inclined respect to the slope $\theta'(s)$. 

Due to mathematical properties of the crack shape (Eq. \ref{theta}), the opening function $\theta(s)$ is even and the maximum $M = \max(\theta'(s))$ is obtained for $s=kp$, then for $s=0$. Then the derivative at $s=0$ belongs to sub-differential defined as $\theta'(0)\in[-M,M]$. The corresponding yield surface is non convex (Fig. \ref{yieldsurf}); it can be seen as the collection of all the possible yield surface for any $s$.\\

\noindent\textbf{Thermodynamics: dissipation}\\

For shake of simplicity we retain an associated model: the potential of dissipation is the elasticity domain. The flow rule is deduced from the normality rule (in which $\dot{\lambda}\geqslant0$) \emph{i.e.} $\dot{s} = \d f / \d S \, \dot{\lambda}= \textrm{sign} (S) \dot{\lambda}$.
%
Then comes that $S$ and $\dot{s}$ have the same sign, proving that the thermal dissipation Eq. \ref{diss1} is positive under the admissibility condition (Eq. \ref{admcoo}). 
\begin{eqnarray}
\dot{\mathcal{D}}&=&S\dot{s}=(B_{0}-\varphi \sigma_{11}) |\dot{s}|\geqslant 0\label{diss1}
\end{eqnarray}

\noindent\textbf{Yield stress in the tension and compression case}\\

The stress is $\sigma=\hat{\sigma}\,\vec{n}\otimes\vec{n}$; with $\vec{n} = \cos(\alpha)\vec{e}_{1}+\sin(\alpha)\vec{e}_{2}$ and (with no restriction) $0\leqslant\alpha\leqslant\pi/2$. In the specimen's orthonormal basis $[\vec{n},\vec{m}]$, the axial strain writes (from \ref{partoche}, \ref{defostruct}):
\begin{eqnarray}
\epsnn &=& \frac{\sigma}{E}+\theta(s) \cos^2(\alpha)+s \sin(\alpha)\cos(\alpha)
\end{eqnarray}

Following in some way \cite{marigo_98}, we consider that all the cracks are \emph{possible} in the material; the one that will open first will be the one whose orientation $\alpha$ leads to the highest value of the yield function. The condition of yielding $f=0$ (Eq. \ref{fcoo}) writes:

\begin{equation}
|\hat{\sigma}| |\theta'(s)\cos^2(\alpha)+\sin(\alpha)\cos(\alpha)| = B_{0}-\varphi\hat{\sigma}\cos^2(\alpha)\label{ftc}
\end{equation}

The positiveness of the second member represents the admissibility condition (Eq. \ref{admcoo}); it is always true in compression but represents a limit in tension. It can be shown that (Eq. \ref{diss1}) yields $\dot{s}\geqslant0$, in tension and $\dot{s}\leqslant0$ in compression. Then, the crack initializes while considering $\theta'(0^+)=M$ in tension and $\theta'(0^-)=-M$ in compression. The crack angle and the stress value is obtained in the two cases while minimizing the (absolute) stress level and we obtain (with superscript $^t$ for tension and $c$ for compression:
\begin{eqnarray}
\sst = \frac{2B_{0}}{\varphi+M+\sqrt{1+(\varphi+M)^2}}&\quad& \tan(2\asst)=\frac{1}{\varphi + M}\\
\ssc = \frac{2B_{0}}{\varphi+M-\sqrt{1+(\varphi+M)^2}}&\quad&\tan(2\assc)=\frac{-1}{\varphi + M}
\end{eqnarray}

The ratio $-\ssc/\sst$ appears to depend only upon the value of $\varphi+M$. Considering a Coulomb's friction $\varphi=0.5$ and $M=1$ (approximatively relevant of measured roughness \cite{schmittbuhl_94}), this ratio is 10.9, typical for concrete \cite{kupfer_69}. The angle of cracks are respectively $\asst=$17 and $\assc=$73 degrees for tension and compression (Fig. 4) and these value are reasonable respect to experiments.\\

\noindent\textbf{Evolution}\\

The post-peak response can then be obtained from (Eq. \ref{ftc}), as angles $\alpha$ are set. The Barenblatt constant $B$ is supposed to depend upon the sliding $s$, in order to represent the damage of the material. Using a linear dependancy, we suppose that no more cohesion remains after an half step:

\begin{equation}
B(s) = B_{0} \left<1- \frac{2 \max_{t}|s|}{p}\right>^+
\end{equation}

The obtained stress to strain curve (for $p=5\,10^{-3}$, $E=70$ GPa, $B_{0}=6$ GPa, Fig. 3) exhibits softening in both tension and compression. The zero stress state is obtained for a larger strain in tension than in compression. These two facts lead to consider that the proposed model is only relevant to yield stress but not yet to describe anelastic behavior.

\begin{table}[htdp]\label{tc}
\begin{center}
\begin{tabular}{cc}
\includegraphics[scale=0.33]{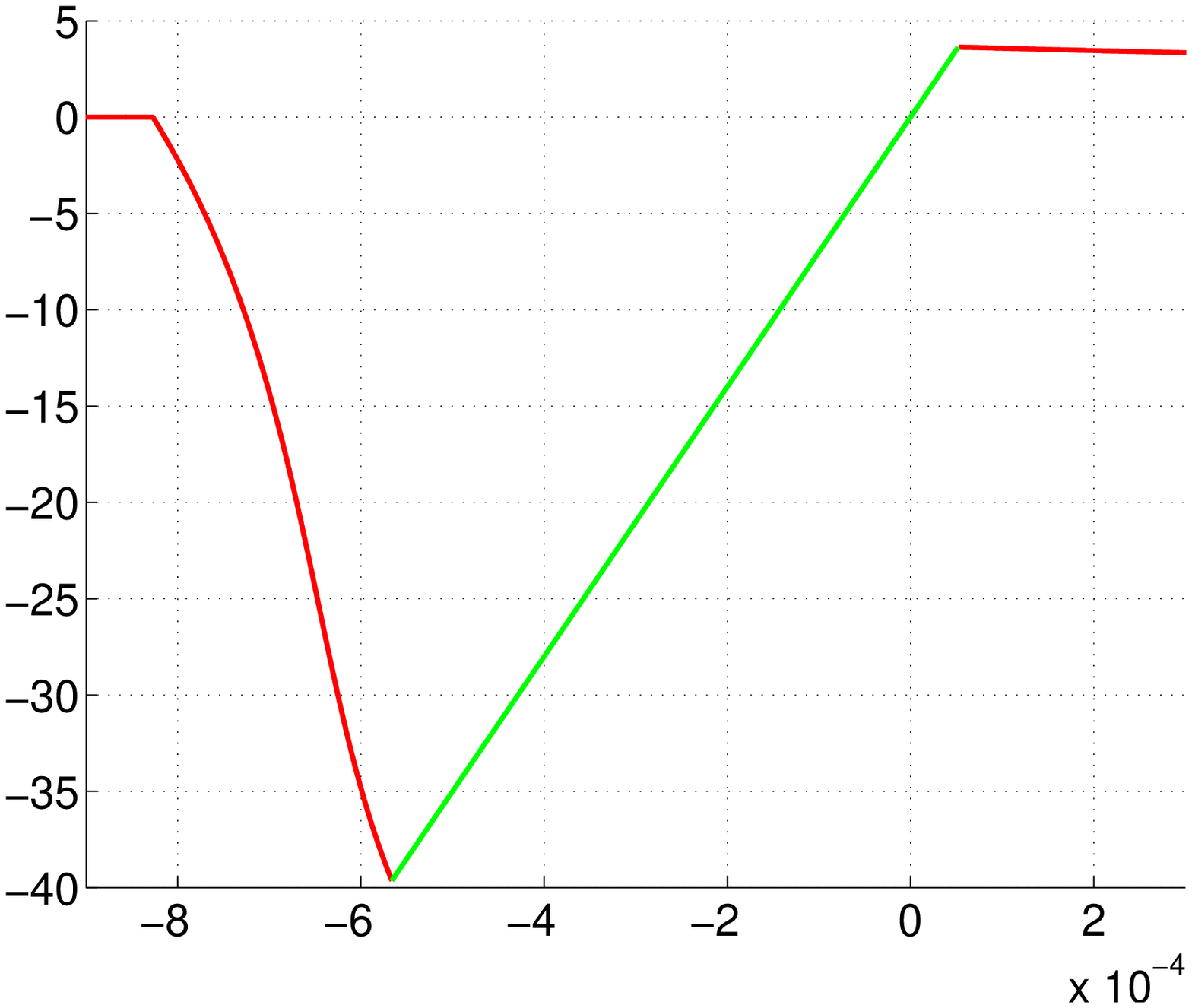}
&
\includegraphics[scale=0.5]{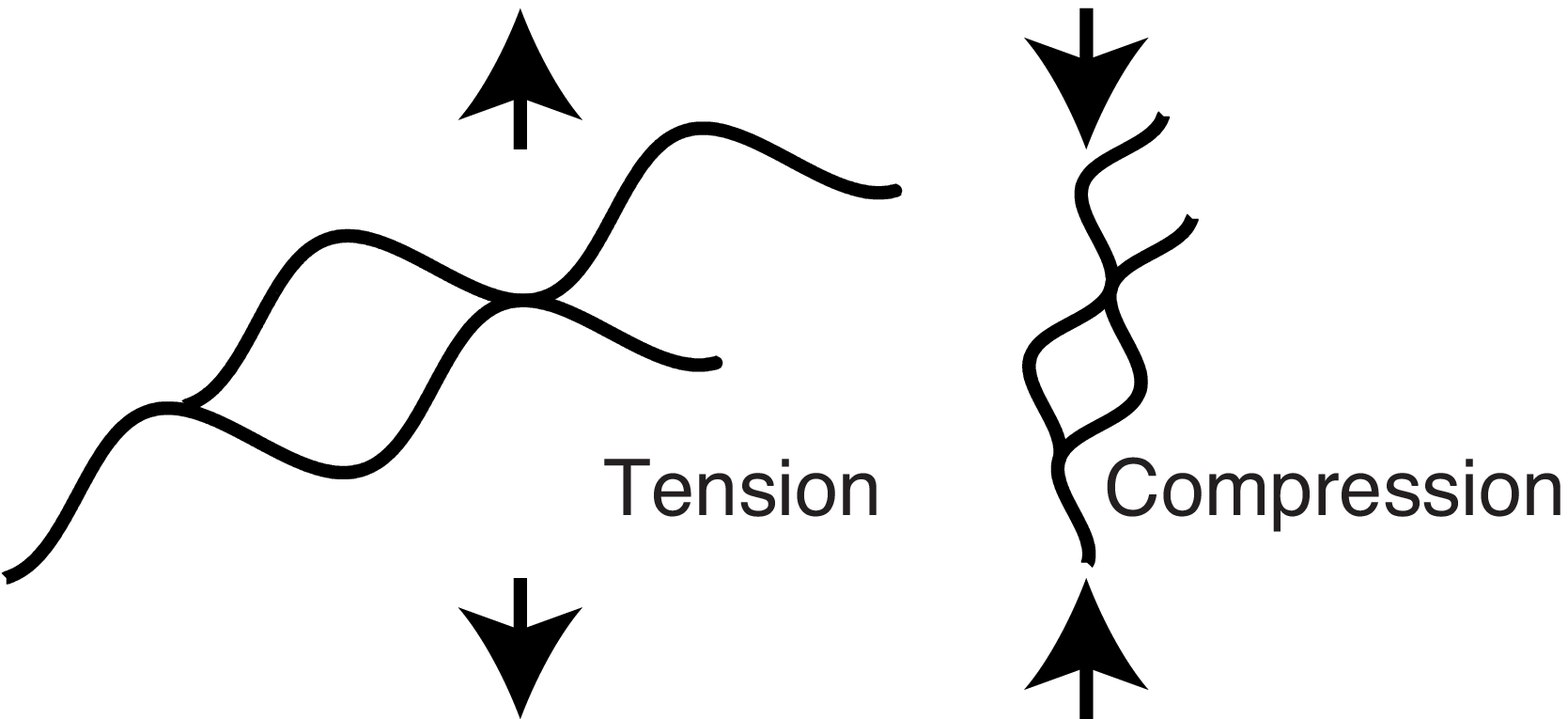}\\
Figure 3: Tension to compression curve & Figure 4: Crack kinematics
\end{tabular}
\end{center}
\end{table}%

\noindent\textbf{Conclusion}\\

This approach seems to be able to predict yield stress and crack angle in tension and compression. An extension in case of bi-tension or compression will be done. The anelastic behavior needs to be enhanced in order to be predictive. Softening behavior will have to be considered with respect to some material scale in order to avoid localization problems (mesh dependancy); this micro-scale being already present in the model allow to think that it is possible.

\bibliography{maderebiblio}

\begin{thebibliography}{1}

\bibitem{delpiero_93}
{\sc G.~Del\_Piero and D.~Owen}, {\em Structured deformations of continua},
  Arch. Rational Mech. Anal., 124 (1993), pp.~99--155.

\bibitem{marigo_98}
{\sc G.~A. Francfort and J.~J. Marigo}, {\em Une approche variationnelle de la
  mécanique du défaut}, ESAIM: Proceedings. Actes du 30 ème Congrès d'Analyse
  Numérique : CANum' 98, 6 (1998), pp.~57--94.

\bibitem{francois_05}
{\sc M.~François and G.~Royer\_Carfagni}, {\em Structured deformation of
  damaged continua with cohesive-frictional sliding rough fractures}, Eur. J.
  of Mech. A/Solids, 24 (2005), pp.~644--660.

\bibitem{kupfer_69}
{\sc H.~Kupfer, H.~K. Hilsdorf, and H.~Rusch}, {\em Behavior of concrete under
  biaxial stresses}, ACI Journal, 66 (1969), pp.~656--666.

\bibitem{schmittbuhl_94}
{\sc J.~Schmittbuhl, S.~Roux, and Y.~Berthaud}, {\em Development of roughness
  in crack propagation}, Europhys. Lett. 28, 28 (1994), pp.~585--590.

\end{thebibliography}

\end{document}